\documentclass[%
 aip, jap
 amsmath,amssymb,
 reprint,%
]{revtex4-1}
\usepackage{graphicx}
\usepackage{dcolumn}
\usepackage{bm}

\usepackage[colorlinks=true,bookmarks=false,citecolor=blue,urlcolor=blue]{hyperref} 
\usepackage{CJKutf8}
\usepackage{amssymb}

\usepackage{txfonts}

\usepackage[utf8]{inputenc}
\usepackage[T1]{fontenc}
\usepackage{etoolbox}
\usepackage{ulem}

\usepackage{xcolor}

\newcommand{\R}[1]{{ \color{red} {\bf} #1}}

\makeatletter
\def\@email#1#2{%
 \endgroup
 \patchcmd{\titleblock@produce}
  {\frontmatter@RRAPformat}
  {\frontmatter@RRAPformat{\produce@RRAP{*#1\href{mailto:#2}{#2}}}\frontmatter@RRAPformat}
  {}{}
}%
\makeatother

\begin{document}
\begin{CJK}{UTF8}{gbsn}

\title{Topological valley crystals in a photonic Su-Schrieffer-Heeger (SSH) variant}
\author{Z. Yu}

\author{H. Lin}%
\email[Electronic mail of corresponding author: ]{linhai@mail.ccnu.edu.cn}
\homepage{http://faculty.ccnu.edu.cn/2006983688}
\author{R. Zhou}
\author{Z. Li}
\affiliation{ 
College of Physics Science and Technology, Central China Normal University, Wuhan 430079, Hubei Province
}%
\author{Z. Mao(毛竹)}

\affiliation{Department of Physics, Hubei University, Wuhan 430062, Hubei Province}

\author{K. Peng(彭旷)}
\affiliation{%
School of Physics and Electronic Science, Hubei University, Wuhan 430062, Hubei Province%
}
\affiliation{Hubei Key Laboratory of Ferroelectric and Piezoelectric Materials and Devices, Hubei University, Wuhan 430062, Hubei Province
}%
\author{Y. Liu(刘泱杰)}
\email[Electronic mail of corresponding author: ]{yangjie@hubu.edu.cn}
\homepage{http://wdxy.hubu.edu.cn/info/1020/1078.htm}
\affiliation{%
School of Physics and Electronic Science, Hubei University, Wuhan 430062, Hubei Province%
}
\affiliation{Lanzhou Center for Theoretical Physics, Key Laboratory of Theoretical Physics of Gansu Province, Lanzhou University, Lanzhou 730000, Gansu Province}
\affiliation{Hubei Key Laboratory of Ferroelectric and Piezoelectric Materials and Devices, Hubei University, Wuhan 430062, Hubei Province
}%
\affiliation{Department of Physics, Hubei University, Wuhan 430062, Hubei Province}
\author{X. Shi}
\affiliation{%
	Wuhan Maritime Communication Research Institute, Wuhan, 430205, China%
}

\date{compiled \today, submitted on 3rd July to J. Appl. Phys., accepted 12nd Sept. 2022. }

\begin{abstract}
	\noindent Progress on two-dimensional materials has shown that valleys, as energy extrema in a hexagonal first Brillouin zone, provide a new degree of freedom for information manipulation. Then valley Hall topological insulators supporting such-polarized edge states on boundaries were set up accordingly. In this paper, a two-dimensional valley crystal composed of six tunable dielectric triangular pillars in each unit cell is proposed in the photonic sense of a deformed Su-Schrieffer-Heeger (SSH) model. We reveal the vortex nature of valley states and establish the selection rules for valley polarized states. Based on the valley topology, a rhombus-shaped beam splitter waveguide is designed to verify the valley-chirality selection rule above. Our numerical results entail that this topologically protected edge states still maintain robust transmission at sharp corners, thus providing a feasible idea for valley photonic devices in THz regime. ~\footnote{JAP22-AR-03586: submitted on 3rd July to J. Appl. Phys., first reports back 1st Aug., revised 14th Aug., and resubmitted 24th Aug., proofed 22nd Sept. 2022. }
\end{abstract}

\maketitle

\end{CJK}

\section{\label{sec:level1}Introduction}

In the past few years, inspired by the development of topological states of matter in condensed matter physics~\cite{Wen2019}, such as quantum Hall states~\cite{Klitzing2017} and topological insulators~\cite{Hasan2010,Qi2011}, the concept of topological band was introduced into valley photonic crystals (VPC) and quickly became a newly-chartered field~\cite{Ozawa2019}. As the wave functions in photonic crystals (PC) appear similar to that of electrons in solid physical systems, various phenomena originally discovered in condensed matter physics, such as quantum Hall~\cite{Klitzing1980}, quantum spin Hall~\cite{Kane2005a,Kane2005b} and quantum valley Hall~\cite{Xiao2007} effect, can be mapped to wave systems in analogue. As pioneers F. Haldane and S. Raghu~\cite{Haldane2008} broke the time-reversal symmetry by applying an external magnetic field to the magneto-optical material, and proposed the photonic quantum Hall effect in PC for the first time. Following the realization of the quantum Hall state using PC~\cite{Joannopoulos1997,Haldane2008,Wang2008,Wang2009}, photonic systems to simulate the quantum spin Hall effect using polarization degeneracy between transverse electric (TE) and transverse magnetic (TM) modes had been extensively studied~\cite{Khanikaev2013,Chen2014,Ma2015,Cheng2016,Slobozhanyuk2019,Bisharat2019,Wu2015}. Furthermore, L.-H. Wu and X. Hu proposed a pair of pseudo-spin states based on inequivalent irreducible representations of the $C_{6v}$ symmetry group, by exploiting the spatial crystal symmetry of a two-dimensional (2D) all-dielectric PC~\cite{Yang2018, Silveirinha2022}. This profound proposal was verified experimentally both in microwave and optical regimes~\cite{Peng2019,Parappurath2020,Liu2020,Zhou2021}. We understand that all these phase transitions derive from the famous Su-Schrieffer-Heeger (SSH) model in condensed matter physics, where one transforms phases via tuning inter/intra-cell coupling between/in unit cells in periodic lattices.

Later on, a new degree of freedom (DOF) from the valley point, was also introduced to the PC platform~\cite{Dong2017,Ma2016,Chen2017,Gao2018,Ma2017,Wu2017,Ye2017,He2019,Shalaev2019,Zhou2022}. Valley points generally emerge at high-symmetry points of the Brillouin zone, and are referred to as minima in the conduction band or maxima in the valence band. The promise of using the valley DOF to store and carry information led to conceptual breakthrough known as valleytronics in electronic applications generally~\cite{Xu2014,Schaibley2016}. Then if the degeneracy between the two valleys is lifted, 2D materials can host the quantum valley Hall effect, which is manifested by a pair of counter-propagating scatteringless states with opposite valley-polarizations at non-trivial domain walls~\cite{Yao2009,Lu2016,Wei2021,Zhang2011,Qiao2011,Zhang2013,Ju2015,Mrudul2021A,Mrudul2021B}. The valley edge state can be realized by combining two types of photonic crystals with different valley Chern numbers. Valley Hall photonic crystals were experimentally demonstrated at various frequencies and have found potential applications in topologically protected refraction~\cite{Zhang2019}, high-efficiency waveguides~\cite{Harari2018,Bandres2018,Noh2020,Zhao2022}, topological waveguide splitters~\cite{Yang2020}, etc. 

In this paper, we propose a 2D VPC in SSH model whose unit cell is composed of six tunable dielectric triangular pillars, which makes use of the reduced symmetry to switch topological phases. First, we propose a topological PC with Dirac points at valley points ($K/{K}^{\prime}$) from its band structures. To verify the topological phase transition, we open the energy band-gap at valleys by breaking the spatial inversion symmetry. By changing the intra/inter-cell coupling strengths, the degeneracy at the Dirac points can be lifted to result in a bandgap. Meanwhile, Berry curvature in the reciprocal space is calculated to show a pair of energy extrema with different signs at the $K/{K}^{\prime}$ point. The topological invariant (i.e., the valley Chern number) of VPCs is calculated by integrating the Berry curvature over the First Brillouin zone (FBZ). Also, we demonstrate that with the valley excited states of our structure, the valley selectivity of PCs is revealed by performing Fourier transform to their wave functions. Finally, we design a beam splitter of rhombus shape to verify the valley-selective transmission. Our work then provides a new idea for THz photonic devices manipulating valley DOF. To note, our model is a 2D variant of photonic SSH\cite{Liu2017,Liu2018} model when we adopt the hexagonal unit cell as shown below.

\begin{figure*}[htbp]
\includegraphics[width=0.9\textwidth]{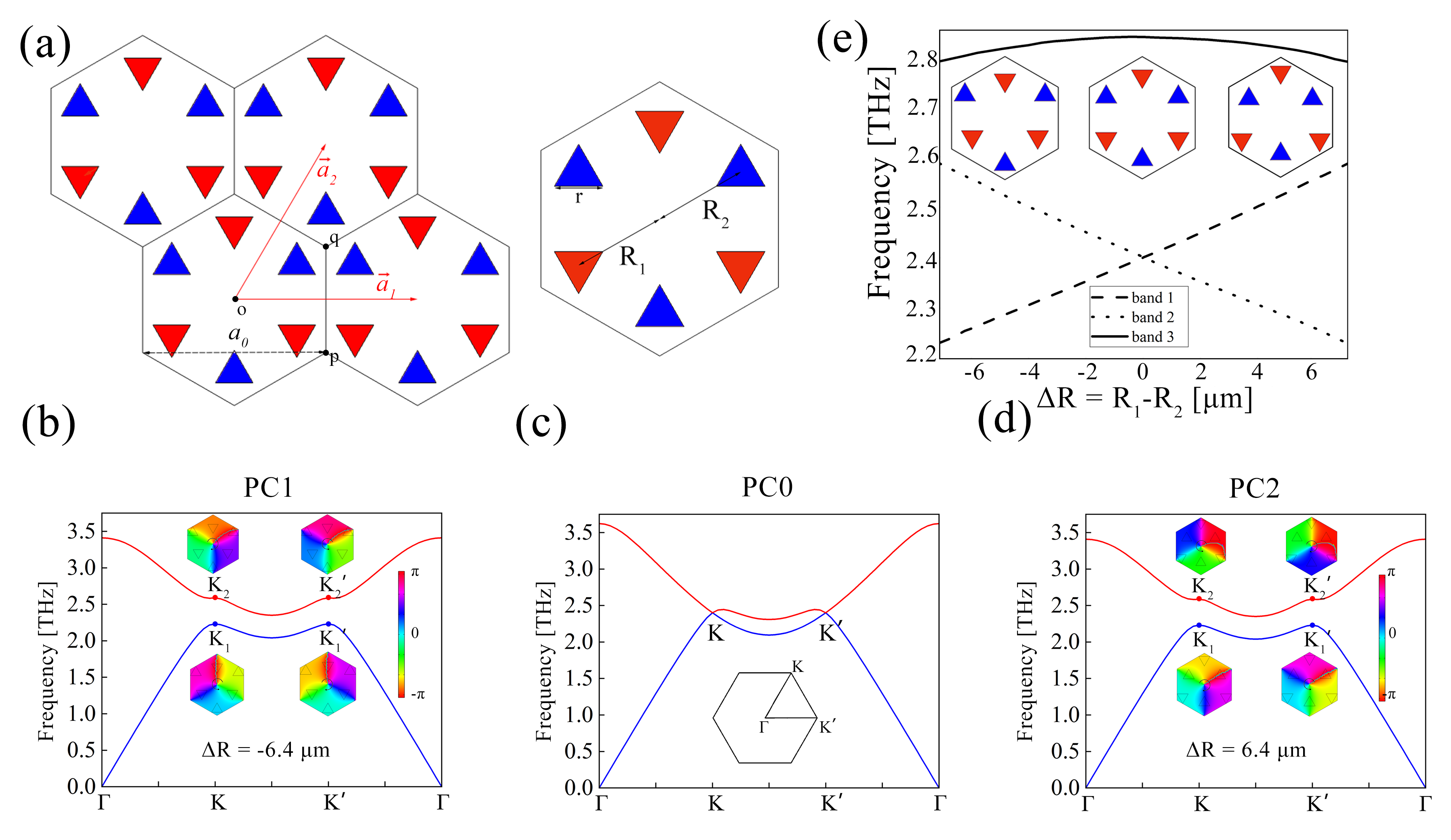}
\caption{\label{fig:epsart} Band structures and topological phase transition. (a) Schematic diagram of a 2D hexagonal lattice with translation vectors of $\boldsymbol{a_1}$ and $\boldsymbol{a_2}$ where the three positions in the lattice are labelled by o, p, q. The lattice constant is $a_0$. Six dielectric pillars with side length of r are placed in the lattice.  (b, c, d) The band structure for the PC with $\Delta R$ = $R_1$ - $R_2$ = -6.4 ${\rm \mu m}$, 0 ${\rm \mu m}$, 6.4 ${\rm \mu m}$ (insets of (b, d) shows the phase distributions). The lower and higher frequency state at $K$/${K}^{\prime}$ are labelled by $K_1$/${K_1}^{\prime}$ and $K_2$/${K_2}^{\prime}$. (b) When $\Delta R$ = -6.4 ${\rm \mu m}$, at the $p_1$ and $q_1$ points the phase distributions reveal that $K_1$/${K_1}^{\prime}$ and $K_2$/${K_2}^{\prime}$ have opposite chirality, whose chirality is indicated by the black arrow in insets. (c) When $\Delta R$ = 0 ${\rm \mu m}$, the Dirac points appear at points $K$ and ${K}^{\prime}$ in the FBZ, and the inset shows the FBZ of the lattice. (d) When $\Delta R$ = 6.4 ${\rm \mu m}$, $K$ and ${K}^{\prime}$ valley points show opposite charility compared to panel (b). (e) Frequency gap variation of PC lattice with change of $\Delta R$ at Brillouin zone center $K$ and the evolution of the unit cell is the inset.} 

\end{figure*}

\section{Numeric model}
In this paper, we consider a 2D hexagonal PC with $C_6$ symmetry as shown in Fig.~1(a), the unit cell of which is composed of six dielectric pillars embedded in air.~\cite{Xi2021,He2019} And the maximal Wyckoff points in the unit cell are represented by labels $o, p, q$ in real space therein. The relative permittivity of the dielectric column is $\epsilon_d = 11.7$, the lattice constant is $a_0 = 50~{\rm \mu m}$, and $\boldsymbol{a}_1$ and $\boldsymbol{a}_2$ are the lattice vectors. In the following, we shall start from the trivial case PC0 and then construct two types of non-trivial cases: PC1 and PC2 respectively. For PC0, the dielectric pillars are equilateral triangles with the side length of $r = a_0/4.9$, and the distance from the center of the lattice to the center of the dielectric column is $R_1$ = $R_2$ = 19.7 ~${\rm \mu m}$ ($R_1$+$R_2$ = 39.4 ~${\rm \mu m}$). We only consider the TM modes and all our results are calculated by \R{the numerical software COMSOL Multiphysics}. The first Brillouin zone (FBZ) in the lattice contains a pair of $K$ and ${K}^{\prime}$ points in its vertices, which are called valley points, as shown in Fig.~1(c) and its inset, where panel (c) displays the band structure of PC0. We use $\Delta R$ = $R_1$-$R_2$ to stand for the intra/inter-cell couplings strength~\cite{XieB2018}. Then PC1 corresponding to $\Delta R = -6.4 ~{\rm \mu m}$, PC0 corresponding to $\Delta R$ = 0 ${\rm \mu m}$, PC2 corresponding to $\Delta R = 6.4 ~{\rm \mu m}$, as shown in panels (b-e) respectively. When $\Delta R$ = 0 [see panel (c)], two degeneracy points appear at $K$/${K}^{\prime}$ point with $f = 2.3 {\rm THz}$ due to the $C_6$ symmetry. By changing $\Delta R$, the intra/inter-cell coupling vary accordingly, and the lattice symmetry is reduced to $C_3$ so that the degeneracy at $K/{K}^{\prime}$ points is lifted, which is shown in panels (b) and (d). For one case of PC1, the bandgap is opened in the frequency range of 2.2-2.35~THz when $\Delta R = -6.4 ~{\rm \mu m}$. For another case, PC2 is constructed by letting $\Delta R = {R_1}-{R_2} = 6.4~{\rm \mu m}$, the corresponding band structure is shown in Fig.~1(d). So for PC2 the degeneracy points at $K/{K}^{\prime}$ are also lifted, and the bandgap appears from ${f} = 2.2\rm~{THz}$ to $f = 2.35 \rm ~{THz}$. We symbol the lower and higher frequency state at valleys by $K_1$/${K_1}^{\prime}$ and $K_2$/${K_2}^{\prime}$ in panels (b) and (d). The valley states in gap exhibit chirality, which is manifest by the phase distribution of $E_z$, i.e., $\arg E_z$. When $\Delta R = -6.4~{\rm \mu m}$ for valley $K$ (PC1), the phases of $K_1$ and $K_2$ have opposite vortex chirality at the position of p and q, and vice versa for ${K}^{\prime}$ valley. And PC2 has the opposite chirality to PC1 at $K/{K}^{\prime}$, indicating a typical band inversion that leads to a topological phase transition. Fig.~1(e) shows the bandgap width corresponding to different $\Delta R$ sizes. The evolution of the unit cell when varying $\Delta R$ is visualized in the Fig.~1(e) inset. When $\Delta R$ changing from $\Delta R = -6.4~{\rm \mu m}$ to $\Delta R = 6.4~{\rm \mu m}$, the bandgap closes and reopens sequentially. This is how we tune the topological phases of VPC by adjusting the structure of unit cells synchronously.

Then we focus on the properties of the $K$-valley state. The top and bottom panels in Fig.~2 represent the phase and amplitude distribution for the periodic lattice, which exhibit the chirality property of valley states. Therein, at the positions of high symmetry, p and q (i.e., threefold rotational symmetry $C_3$), $\left|E_z\right|$ vanishes and the phase goes singular for both $K_1$ and $K_2$ valleys. The black arrows in the lower panel represent time-averaged Poynting vectors $\mathbf{S}=\Re\left(\mathbf{E} \times \mathbf{H}^{*}\right) / 2$, which reveals a typical feature of the vortex field. Therefore, we can control the chirality of the valley vortex by switching the source chirality induced from positioning the dielectric pillars~\cite{XuL2016, Chen2017, Gao2018, Zhou2022}. 

\begin{figure}
\includegraphics[width=0.48\textwidth]{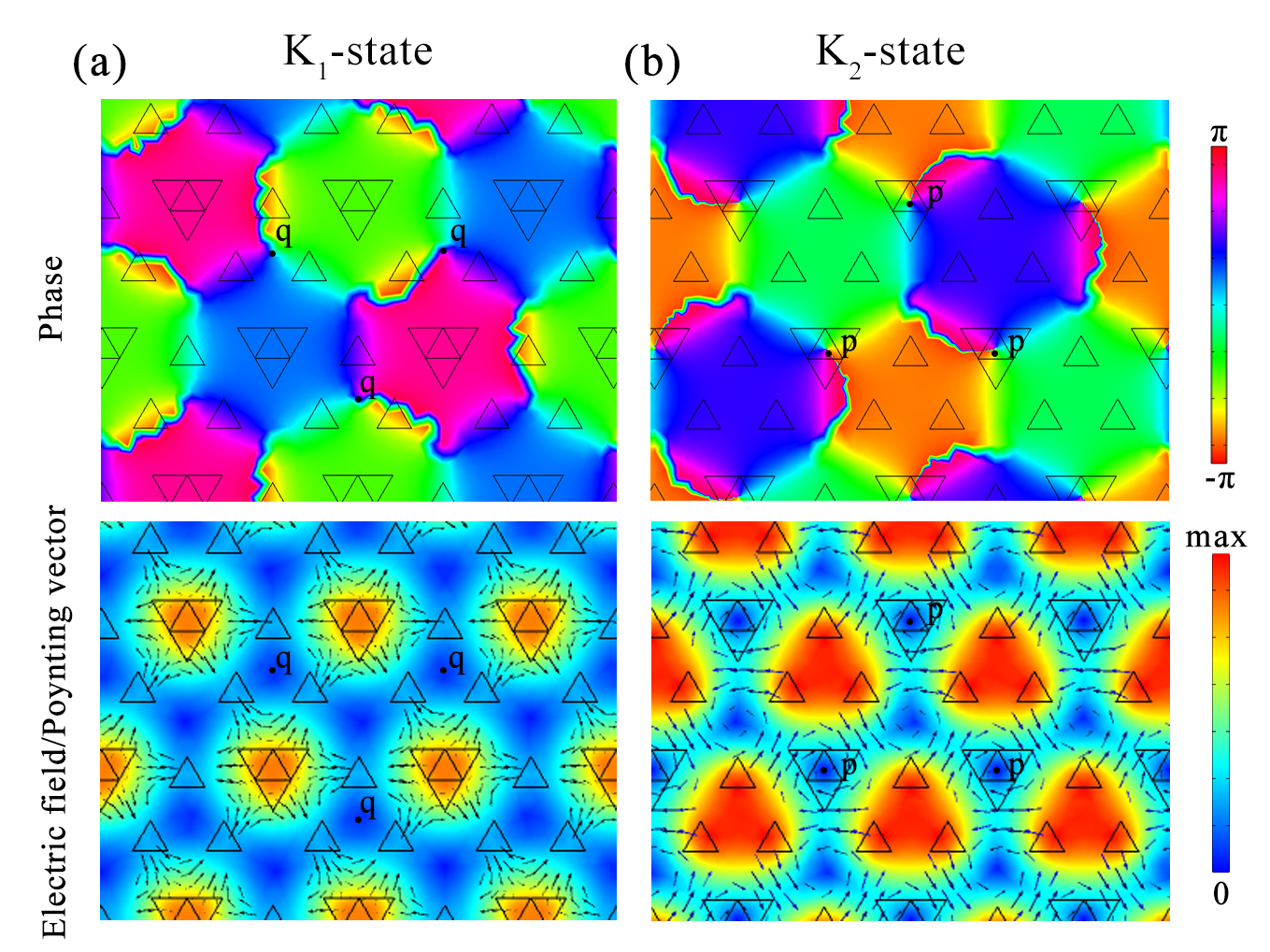}
\caption{\label{fig2}Electric field distribution $\left|E_z\right|$ of the K-valley state [(a) low frequency $K_1$,  (b) high frequency $K_2$] at positions $p, q$. The upper and lower panels respectively represent the valley phase and electric field amplitude of the periodic lattice, and the arrows in the lower panel indicate the corresponding time-averaged Poynting vector.}
\end{figure}

We then verify the valley selectivity~\cite{Settnes2016} of the structure in Fig.~3. Therein a chiral source with $m = \pm1$ is placed at the center of the eddy current in $K$/${K}^{\prime}$ state, of which $m$ is the vortex index of the field $E_z$. And the excited electric field $E_z$ distribution is shown in the left panel. The spatial Fourier spectrum is plotted in the right panel of Fig.~3 where the green solid hexagons represent the FBZ. These results show that ${K}^{\prime}$ state is excited when the source goes $m = -1$, and the $K$ state when $m =+1$.

\begin{figure}
\includegraphics[width=0.48\textwidth]{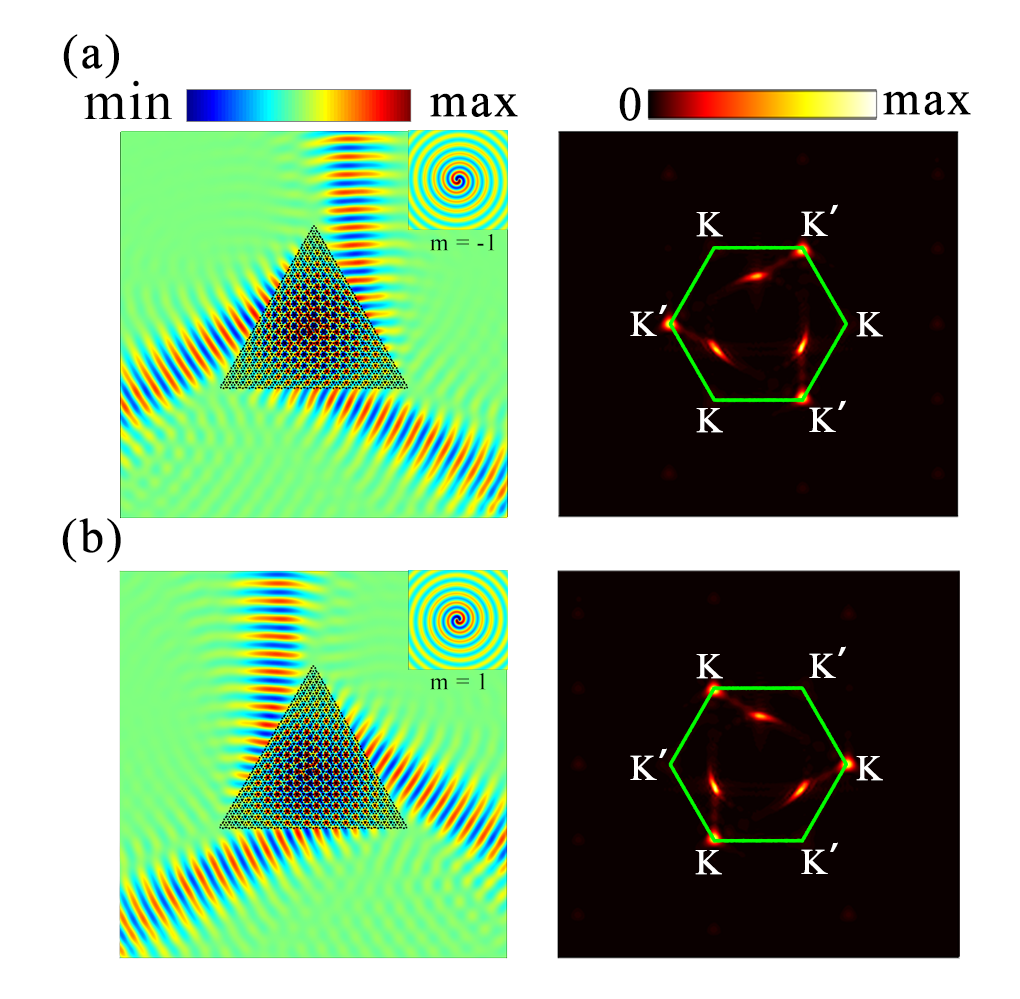} 

\caption{\label{fig3} Electric field $E_z$ stimulated by a chiral source with (a)$m = -1$ and (b) $m = 1$ positioned in the center of samples. The insets in the left panels represent the vortex feature of the chiral source.  The right panels display the corresponding Fourier spectra in the momentum space, where the green lines mark the boundary of the hexagonal FBZ.}
\end{figure}

To verify the topological phase transitions, non-zero Berry curvature and valley Chern number in the inversion-asymmetric valley PC are calculated in this section. We construct a minimal band model of bulk dispersions and derive the photonic effective Hamiltonian based on the $\mathbf{k}\cdot\mathbf{p}$ approximation\cite{Ma2016,Chen2017}

\begin{equation}
	\hat{H}=v_D\left({\hat{\sigma}}_x{\hat{\tau}}_z\delta k_x+{\hat{\sigma}}_y\delta k_y\right)+\lambda_{\varepsilon_z}^P{\hat{\sigma}}_z,
\end{equation}
where $\delta\vec{k}$ measures displacement from the valley center ${K}^{\prime}$ or $K$ point. ${\hat{\sigma}}_i$ and ${\hat{\tau}}_i$ are Pauli matrices acting on sublattice and valley spaces, respectively. $\lambda_{\varepsilon_z}^P{\hat{\sigma}}_z$ indicates a frequency band gap due to the inversion asymmetry. And $\lambda_{\varepsilon_{z}}^{P} \propto\left[\int_{\mathrm{B}} \varepsilon_{z} \mathrm{~d} s-\int_{\mathrm{A}} \varepsilon_{z} \mathrm{~d} s\right]$ denotes the integration of $\varepsilon_z$ at the dielectric rod regions. In Fig.~4, another manner to look at the same PC via a differently-defined unit cell is shown, and the dashed rhomboid denotes a unit cell of PC. When $\Delta R = -6.4 ~{\rm \mu m}$, as shown in Fig.~4(b), $\int_{\mathrm{B}} \varepsilon_{Z} \mathrm{~d} s<\int_{\mathrm{A}} \varepsilon_{Z} \mathrm{~d} s$ leads to $\lambda_{\varepsilon_z}^P<0$ and a complete band gap appears and the other way round in Fig. ~4(d). In Fig.~4(c), $\Delta R = 0 ~{\rm \mu m}$, $\int_{\mathrm{B}} \varepsilon_{Z} \mathrm{~d} s=\int_{\mathrm{A}} \varepsilon_{Z} \mathrm{~d} s$ and the band gap disappear with a degeneracy point is located in $K/{K}^{\prime}$ points. Moreover, the effective Hamiltonian implies a valley-dependent topological index of Berry curvature. The Berry connection of the lowest band can be defined as
\begin{equation}
\vec{A}(\vec{k})\equiv i\left\langle u_{\vec{k}}\left|\nabla_{\vec{k}}\right|u_{\vec{k}}\right\rangle,
\end{equation}
where $\vert u_{\vec{k}}\rangle$ is the electromagnetic field, an asterisk denotes complex conjugation, and $\epsilon(\vec{r})$ is the spatial permittivity. And Berry curvature $\Omega(\vec{k})$ is
\begin{equation}
\Omega(\vec{k})\equiv\nabla_{\vec{k}}\times\vec{A}(\vec{k})=\frac{\partial A_y(\vec{k})}{\partial k_x}-\frac{\partial A_x(\vec{k})}{\partial k_y}. 
\end{equation}
The topological features of VPCs are related with the Berry curvature in the FBZ. As shown in Fig.~5(a), Berry curvature shows opposite signs at the $K$/${K}^{\prime}$ points. For PC1 Berry curvature distribution around $K$ are negative in value, and that around $K'$ positive; and for PC2 vice versa. Therefore, the integration of Berry curvature over the whole FBZ is zero. Topological indices at $K$ and ${K}^{\prime}$ valleys, defined as the integration of Berry curvature within half of the Brillouin zone(HBZ), can be calculated as:
\begin{equation}
C^{K / K^{\prime}}=\frac{1}{2 \pi} \int_{\text{HBZ}} \Omega(k) d^{2} k=\pm \frac{1}{2} {sgn}\left(\Delta_{p}\right). 
\end{equation}

\begin{figure}
	\includegraphics[width=0.48\textwidth]{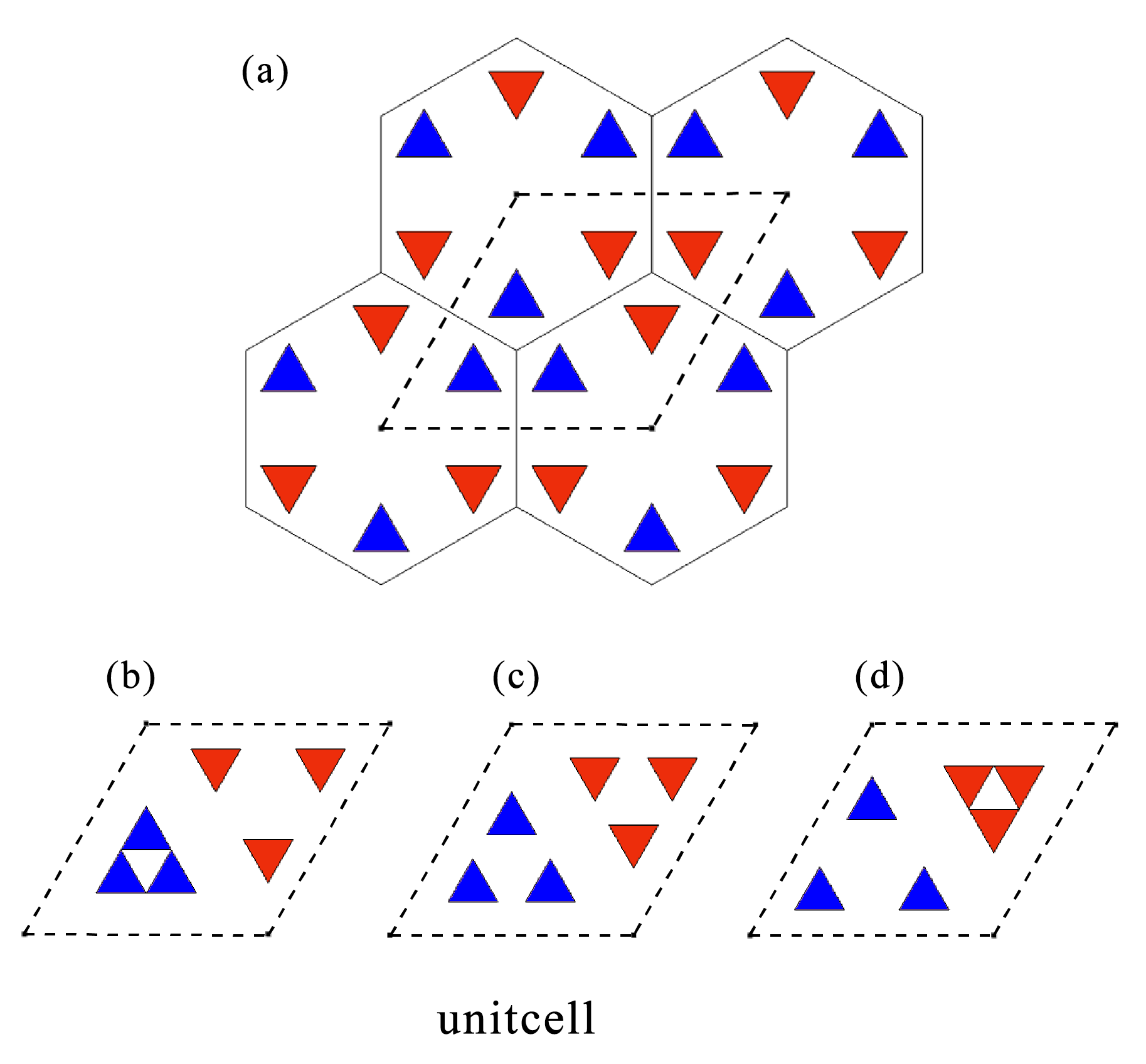} 
	
	\caption{\label{fig4} Schematic diagram of a 2D hexagonal lattice of dielectric rods embedded in an air background. e.g., blue and red rods in the dashed rhomboid. (b), (c) and (d) represent the unit cell with $\Delta R = -6.4~{\rm \mu m},0~{\rm \mu m}$  and  $6.4 ~{\rm \mu m}$.}
\end{figure}

The valley Chern number is only determined by the sign of $\Delta_p$. And the nonzero topological invariant, i.e., valley Chern number is $C_V=\left(C_K-C_{K\prime}\right)$~\cite{Chen2017,Phan2021}. As depicted in Fig. ~5(b), when $\mathrm{\Delta R}<0$,  $C_V=-1$, and for $\mathrm{\Delta R}>0$, $C_V=1$. Therefore PC1 and PC2 have quantized valley Chern number with opposite signs, which leads to a topological phase transition. 

\begin{figure}
\includegraphics[width=0.48\textwidth]{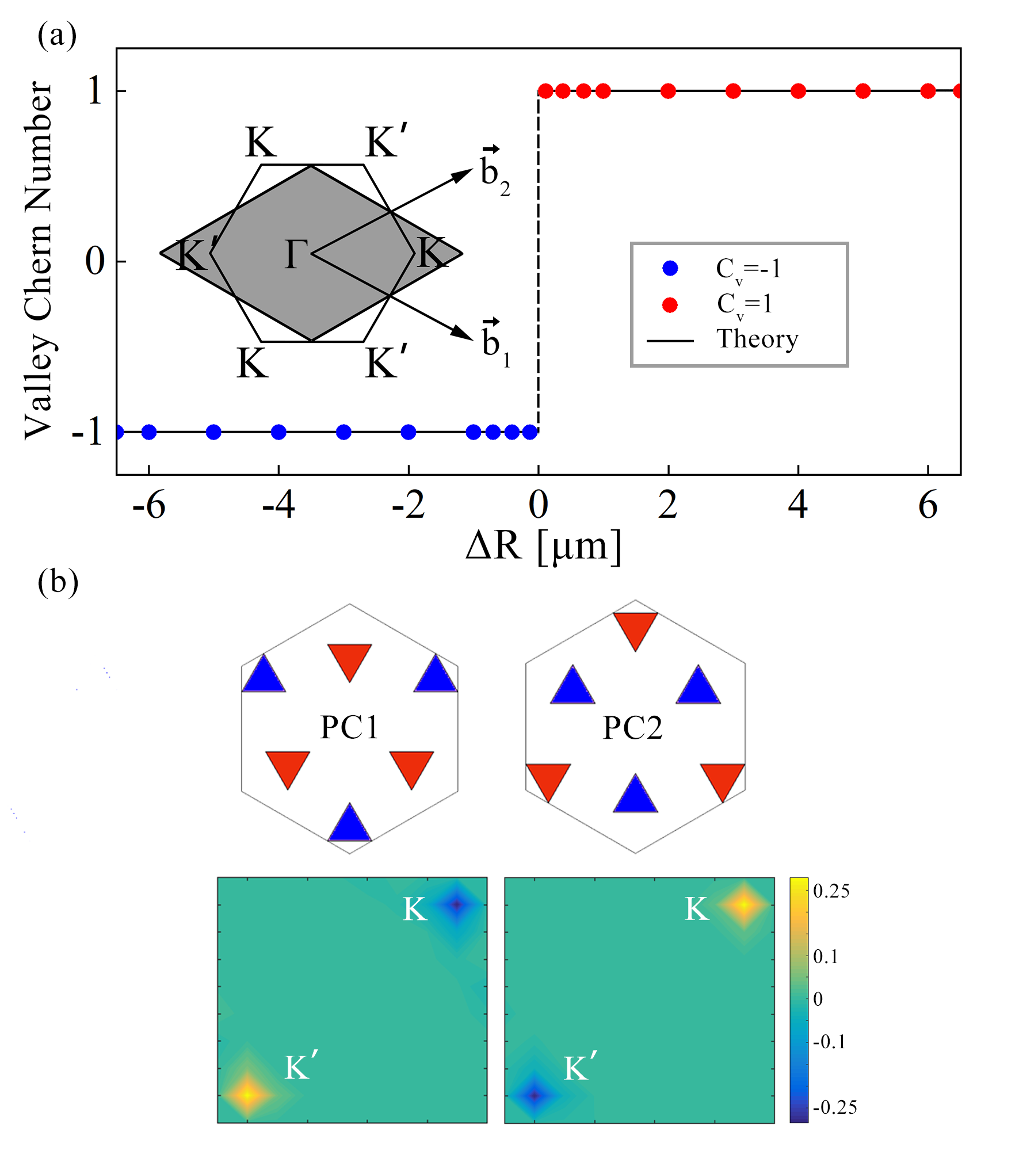}
\caption{\label{fig5} Valley Chern numbers. (a) Two PC structures and their corresponding Berry curvature distribution in the FBZ.  (b) The variation of valley Chern number with change of $\Delta R$. Blue dots for $C_v=-1$, red dots for $C_v=1$, and the black line for theoretical calculation with FBZ shown as the inset.}
\end{figure}

\section{RESULTS AND DISCUSSION}

\textbf{Topological edge states at the zigzag interface}: As the sign of the difference of the valley Chern numbers determines the propagation direction of the emerging edge states, the edge states at the two valleys have opposite velocities, locked by the valley states. To confirm this numerically, Fig.~6(a) displays a $30\times1$cell structure composed of PC1 and PC2 constructed in our simulation. The separation interface is made of 30 cells in y-direction and infinite in x-direction. In simulation, Floquet periodic boundary conditions are imposed on the left and right boundaries of the supercell, and the upper and lower boundaries satisfy scattering boundary condition. Furthermore, Fig.~6.(b) shows the spatial distribution of electric field with the frequency of 2.35~THz at $k_x$=$\pm$0.35$\times2\pi/a_0$. And the energy band for our valley crystals is shown in Fig.~6(c), where the red curve represents the edge state with topological protection while the gray region represents the bulk band. And the black arrows represent the the Poynting vectors. It can be seen that the group velocities of the edge states at different valleys are opposite, indicating the valley-momentum locking behavior. Owing to the presence of nontrivial valley Chern number, the localized EM states are observed along the interfaces. Since the topological edge states are locked within the $K/{K}^{\prime}$ valleys, inter-valley scattering is strongly suppressed despite the presence of obstacles of sharp corners. Such properties make the designed VPC an excellent candidate as waveguides.

\begin{figure*}[htbp]
\includegraphics[width=0.96\textwidth]{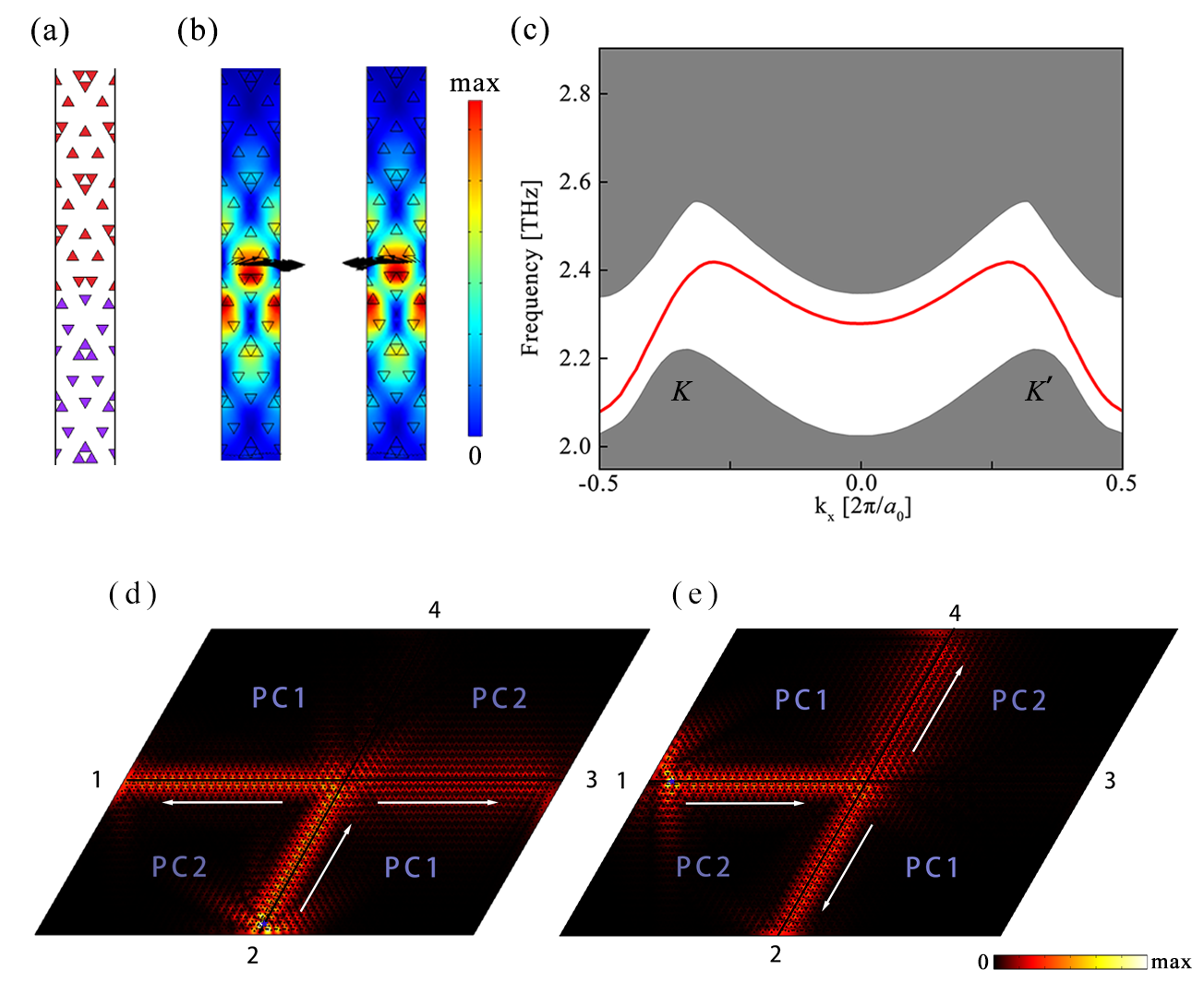} 

\caption{\label{fig6} A supercell consisting of PC1 and PC2 with zigzag interfaces and topological transport of valley-polarized edge states. (a) Schematic of a supercell consisting of PC1 and PC2 with zigzag interfaces. (b) The electric field distribution with the frequency of 2.35 THz at $k_x$ = $\pm0.35\times2\pi/a_0$. And the black arrows represent the average power flow (i.e., Poynting vectors) (c) Band dispersion of the edge and bulk states. (d-e) Simulated field maps of the electric field at 2.26 THz on $xy$ plane with a chiral source at (d) port 1 and (e) port 2, as a rhombus-shape splitter that consists of PC1 and PC2, where the excited valley-dependent edge state by a chirality source (marked by blue asterisks). }
\end{figure*}

\textbf{Verification of valley-selective transmission}: The valley-momentum locking edge states discussed above aim for designing functional EM devices, and henceforth we designed a rhombus-shaped beamsplitter (shown in Fig.~6) for prototype. The beamsplitter consists of four regions, which is selectively activated by the valley degree of freedom. The upper-left and lower-right areas are filled by PC1, and the lower-left and upper-right areas are filled by PC2. In this way PC1/PC2 and PC2/PC1 interface domain walls between distinct PCs. By placing a chiral source, we calculate the electric field in $xy$ plane of the diamond waveguide to directly visualize the edge states and to demonstrate the valley polarization. The simulated electric fields are shown in Fig.~6(d) and (e), where the blue asterisk is the source and the white arrow the direction of the propagation edge state. At $f = 2.26 \rm ~THz$, the edge state is successfully excited by the chiral source and the electric field is well confined along the interface and propagates only in the direction correlating with the chirality of the source. Fig. ~6(d) shows that when the chiral source is excited at the port 2, the wave propagates to port 1 and port 3. Whereas it is excited from port 1 shown in the Fig. ~6(e), wave travels to the port 2 and port 4 barely leaking to port 3. This routing result is guaranteed by the different valley pseudospin of the interface states.

\section{\label{sec:level3}CONCLUSIONS}
In conclusion, we design a VPC to reveal the dynamic process of topological phase transition by turning the inter/intra-cell coupling strength. In our design, a VPC based on broken $C_6$ symmetry is adopted with Dirac points at $K/{K}^{\prime}$ points, and by twisting the dielectric columns in unit cells the intra/inter couplings are adjusted leading to a reduced symmetry of $C_3$, when the degeneracy at Dirac points is lifted. Then our two distinct valley states are demonstrated to verify the vortex feature of the wave function. Also we calculate the valley Chern number of the VPC accordingly and reveal the topological phase transition. Finally, we confirm the valley selection principle of the designed VPC in a beam splitter of rhombus shape. Our work then provides a new experimental setup for application of THz VPC devices.  

\appendix
\section{More on the periodic unit lengths setup in numerics for supercells}

\begin{figure*}[htbp]
	\includegraphics[width=0.96\textwidth]{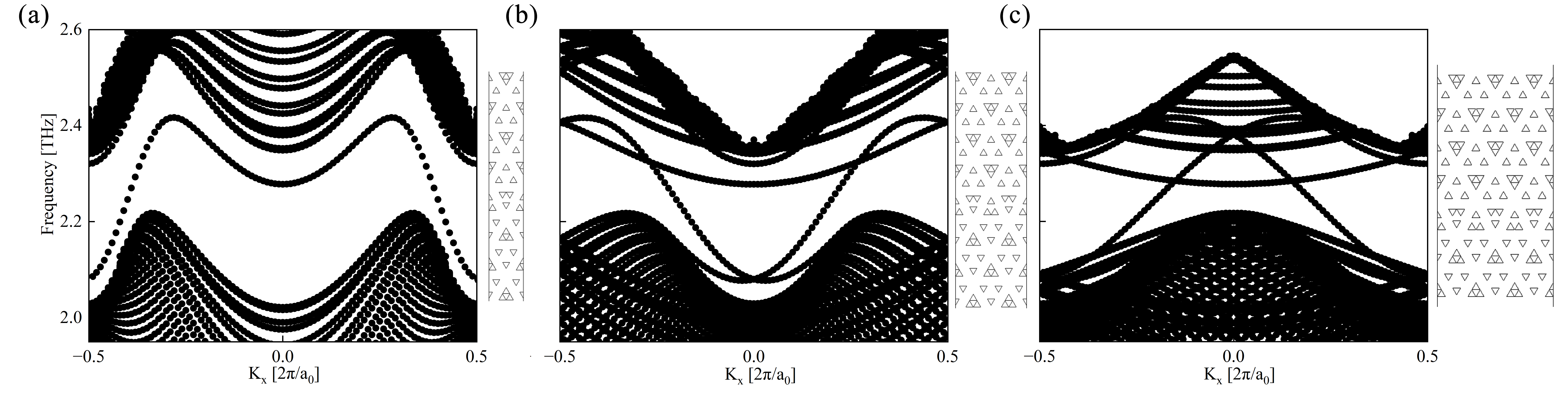} 	
	\caption{\label{fig7} Three cases for dispersion bands of a VPC serving as building blocks of our beam splitter. Band structure of (a) $30\times1$ (b)$30\times2$ (c)$30\times3$ supercell and the schematic of the supercell composed of PC1 and PC2 with zigzag interfaces is on the right of each dispersion band. }
\end{figure*}

In Append.~A, we plot three cases in Fig.~7 for dispersion bands of a VPC serving as building blocks of our beam splitter in Fig.~6. This means that unit cell setup in one of the two dimensions may not be the only feasible one. 

\section{Field distributions on more working frequencies of beam-splitters}

\begin{figure*}[htbp]
	\includegraphics[width=0.96\textwidth]{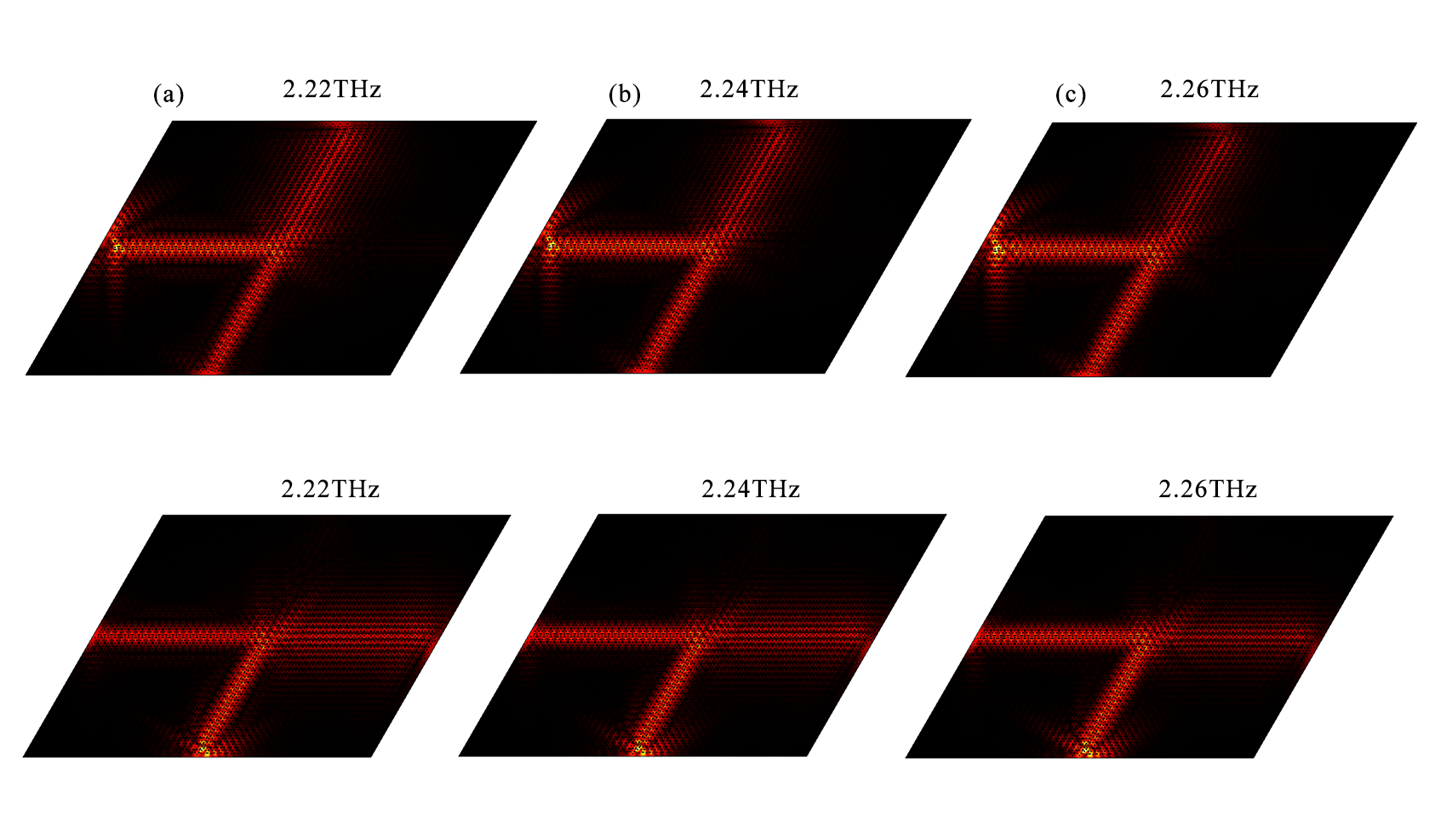} 	
	\caption{\label{fig8} Electric field distributions in simulation at (a) 2.22 THz, (b) 2.24 THz and (c) 2.26 THz on the xy plane with a chiral source at port1 (lower panel) and port 2 (upper panel). The setup is the same as Fig.~\ref{fig6} (d-e).  }
\end{figure*}

In Append.~B, we simulate the electric field of the beam-splitter on more frequencies in the gap. As seen in Fig.~8, at  $f = 2.22{\rm THz}, 2.24{\rm THz}$ and $2.26{\rm THz}$, the edge states are successfully excited by the chiral source.

\begin{acknowledgments}
H. L. and Y. L. acknowledge helpful guidance and support from Zhou B. and Xu D.-H., and Y. L. thanks Zhao Y. for helpful discussion. Z. Y., H. L. and R. Z. thank Fundamental Research Funds for the Central Universities of China (CCNU20GF004, CCNU19TS073); Open fund of China Ship Development and Design Centre under Grant (XM0120190196); Guangxi Key Laboratory of Wireless Wideband Communication and Signal Processing, Guilin University of
Electronic Technology (GXKL06190202) and Hongque Innovation Center (HQ202104001). H. L., Z. M., K. P. and Y. L. are supported by 
Young Scientists Fund (NSFC11804087); National Natural Science Foundation of China (NSFC41974195, NSFC12047501); Science and Technology Department of Hubei Province (2018CFB148,2020CFB266); Department of Education in Hubei Province (T2020001-030200301301002, Q20201006, Q20211008). 
\end{acknowledgments}

\section*{Data Availability Statement}

The data that support the findings of this study are available from the corresponding authors upon reasonable request. 

\section*{Author contribution}
Z. Y. and H. L. proposed the idea. Z. Y. performed the calculation, produced all the figures, and wrote the manuscript draft. R. Z. and Z. L. contributed to the calculation tools. Z. Y. and Y. L. analyzed the data. H. L. and Y. L. lead the project and revised the whole manuscript thoroughly. Z. M. , K. P. and X. Shi contributed to analyzing the data and to revising the paper.

\nocite{*}
\bibliography{ref2}

\end{document}